# Gravity inversion using a directional filtering method

Orlando Silva[1]     Pedro Nogueira[2]

[1] Physics Dep., Évora University, Portugal.     Email: orlando@uevora.pt

[2] Geosciences Dep., Évora University, Portugal.     Email: pmn@uevora.pt

**Abstract**

The calculation of the underground density field from measured gravity data has been done by a variety of methods of varied types. The use of the vector gravity components is here addressed in order to develop one accurate gravity inversion method. The equation to directly calculate the Cartesian components of the gravity field from the density field is transformed in one non-exact correction equation. The discrete vector equations are transformed into scalar equations by using one vector function as directional filter. The used equations have the property that the convergence of the calculated density field to the exact values may only happen if the same occurs to the gravity field. The equations of the method so admit asymptotic exact solutions to the gravity inversion problem. The use of a smooth directional filter allows the equilibrated influence of all the gravity data over all the calculated density deviation corrections. The unknown density field is calculated by means of one iterative under-relaxed method. Two synthetic gravity fields are inverted by means of the presented method. The calculated shallow density fields are very accurate while the far density fields apparently converge to the correct values at very low convergence rates.

**Keywords:** gravity inversion • directional filtering method • accurate solutions

## 1. Introduction

Disregarding the gravitational and dynamic effects of mass distributions external to the Earth, the gravitational field may be calculated by Eq. (1)

$$\boldsymbol{g} = -G \int_{V_E} \frac{\rho dV}{r^2} \boldsymbol{r}_u \qquad (1)$$

where $G$ is the universal gravity constant, $V_E$ is the volume domain of Earth and $\boldsymbol{r}_u$ is the unity vector of the direction from mass to the point where $\boldsymbol{g}$ is calculated.

In order to determine the gravity effects of an Earth volume subdomain V, Eq. (1) is decoupled in two non-intersecting integrals,

$$\boldsymbol{g} = -G \left( \int_{V_E-V} \frac{\rho dV}{r^2} \boldsymbol{r}_u + \int_V \frac{\rho dV}{r^2} \boldsymbol{r}_u \right) \qquad (2)$$

The gravity field in the region enclosing V is simplified by defining one reference regional density $\rho_{ref}$ that contributes to the coarse regional gravity field $\boldsymbol{g}_{ref}$. Any important irregularities of the local gravity field will then be attributed to the effects of the density distribution inside V. Replacing the reference regional fields in the first term of right hand of Eq. (2) and subtracting $\rho_{ref}$, the gravity deviations are calculated from density deviations $\delta\rho = \rho - \rho_{ref}$ inside V in the way expressed by Eq. (3).

$$\boldsymbol{\delta g} \equiv \boldsymbol{g} - \boldsymbol{g}_{ref} = -G \int_V \frac{(\rho - \rho_{ref}) dV}{r^2} \boldsymbol{r}_u \qquad (3)$$

Regular and realistic fields $\boldsymbol{g}_{ref}$ and $\rho_{ref}$ must be found from regional geological and geophysical data, possibly by simply interpolation and extrapolation of the available measured values of $\boldsymbol{g}$ and $\rho$ in chosen domains enclosing V. Assuming the validity of Eq. (3), gravity fields and their deviations can be directly calculated from Eq.s (1, 3) in terms of the density and density deviation fields, once reference fields have been uniquely ascribed.

Unlike the calculation of gravity deviations from the density deviations by Eq. (3), the calculation of the density deviations from the gravity deviations has no simple expression with known parameters, except in very simple cases. A great number of varied inversion methods have been built, e.g. Green (1975), Mosegaard (1995), Li (1998), Zhdanov (2004), Shamsipour (2010), Camacho (2011), and references therein.

The system composed by density and gravity data, defined in their respective spatial domains, supplied with the Newton gravity rules, will be modelled by means of discrete variables to which the same gravity law applies. The simple calculation of the Cartesian components of the gravity field from the density field will be inverted by means of one method that iteratively calculates density from gravity data.

In the present work the density deviation $\delta\rho$ will be obtained from hypothetically error-free gravity deviation $\boldsymbol{\delta g}$ vector components after discretization of Eq. (3). The resulting algebraic equations will then be modified to generate scalar equations. The inversion method using all directional gravity data is intended to reach high accuracy in the obtained density field inside V. This expectation is related to the use of the supposedly exact gravity vector field in a way in which all the density field is obtained from enough gravity data.

The summation of the discretized equations over all the gravity points generates equations that link each density value to all gravity values. These equations are then split in one exact and one inexact equations. Their association gives rise to approximated equations to calculate density corrections from gravity corrections.

The transformation of the vector equations in scalar equations is executed by applying smooth directional filtering to each correction of the deviation gravity vector. The effects of the gravity correction vectors on the density corrections depend on the directions of gravity corrections.

The use of vectorial gravity fields in the inversion process is expected to improve the accuracy of the results, as a consequence of the directional relation between density and gravity locations. When no measured vector $\boldsymbol{g}$ field is available, the transformation of gradiometry data, e.g. Zhdanov (2004), into gravity field may be effected.

Trying to avoid undetermined or impossible solutions, the number of gravity data is made equal to the density data number. This can be done, in case of scarce gravity data, by means of non-linear interpolations and extrapolations. Exact $\delta\rho$ results are expected under the theoretical supposition of sufficient and exact $\boldsymbol{\delta g}$ data. In cases in which only scalar data are available, new versions of the method are required.

As distance increases, the decreasing effect of mass over gravity leads to smaller sensitivity of gravity to distant mass distributions. The inverse problem leads to large density errors in the calculation of the deep density field. The weakening link between gravity and density as distance increases may cause almost indistinguishable gravity fields caused by considerably different density fields.

The gravity inversion method presented will be applied to synthetic gravity fields obtained from arbitrarily chosen density fields.

## 2. Inversion method

An approximation to the continuous Eq. (3) is the discretized set of Eq.s (4) linking N gravity deviations to the M density deviations within V. This equation will be used as basis to develop the inversion method.

$$\delta g_n \equiv g_n - g_{ref_n} = \sum_m \alpha_{nm}(\rho_m - \rho_{ref_m}) \equiv \sum_m \alpha_{nm}\delta\rho_m \quad ; \quad \alpha_{nm} = -G\frac{\delta V_m}{r_{mn}^2}\cdot r_{u_{mn}} \tag{4}$$

where $r_{u_{mn}} = r_{mn}/|r_{mn}|$, $\delta V_m$ are small volumes centered at $r_m$ where density deviations $\delta\rho_m$ are defined, while $\delta g$ field is defined at points $r_n$ or small areas not necessarily contiguous.

The use of density variables must not hide the fact that all gravity and inversion calculations in the present study use $\alpha_{nm}$ defined in terms of locations $r_n$ and $r_m$ defined at the grid centers. More refined grids or analytic calculations of $\alpha$ would better represent the effects of real density fields on gravity. When $\delta V_m$ and the areas cannot be considered small, analytical or numerically calculated corrections to $\alpha_{nm}$ may be used for enhanced precision.

Presumed exact gravity data $g_n^l$ can be obtained by presumed exact measurements or calculated from presumed exact $\rho_m^l$ density values. Correspondingly, possibly inexact $g_n^c$ can be obtained from possibly inexact $\rho_m^c$.

Eq. (4) can be used to calculate density fields on the basis of presumed exact gravity data $g_n^l$. The modification of only one exact $\delta\rho_m^l$ to one inexact $\delta\rho_m^c$ will modify all the $\delta g_n^l$ to $\delta g_n^c$ such that

$$\delta g_n^l - \delta g_n^c = \alpha_{nm}(\delta\rho_m^l - \delta\rho_m^c) \tag{5}$$

Eq. (5) links the correction of one sole density deviation to corrections of all the gravity deviations, expounding the fact that the corrections of gravity deviations have the directions of $\alpha_{nm}$. One equally valid equation would result by multiplying both sides by the same factor.

Exact gravity deviations are presumed to be known. If only one density deviation has the non-exact value $\delta\rho_m^c$, each single Eq. (5) can be used to calculate the exact $\delta\rho_m^l$, while $\delta g_n^c$ is calculated by means of Eq. (4).

In the case in which various density deviations are not exact, Eq. (5) ceases to be correct, so invalidating the possibility of directly finding one density correction from the gravity corrections. Indirect procedures are then needed so that exact density deviations can be determined from the set of gravity deviations.

All gravity and density deviations are linked so that the density deviations field must then be calculated from the ensemble of all the gravity deviations.

In order to calculate density corrections from all gravity data, Eq. (5) is summed over all the gravity data. Moreover, to filter the effects of all gravity corrections in the directions of $r_{mn}$, one multiplier coefficient is introduced in that direction so leading to Eq. (6).

$$\sum_n f_{nm}\cdot(\delta g_n^l - \delta g_n^c) = \sum_n (\alpha_{nm}\cdot f_{nm})\cdot(\delta\rho_m^l - \delta\rho_m^c) \tag{6}$$

If there is only one non-zero $\delta\rho_m^l - \delta\rho_m^c$, either Eq.s (6) or (5) can be used to calculate all the corrections to the gravity deviations. If however density corrections at different locations are responsible for the gravity corrections, Eq. (6) ceases to be equivalent to Eq. (5) and both of them will usually be false. The importance of Eq. (6) is founded however on the possibility of using $f_{nm}$ as one filtering device to carry out one under-relaxed iterative calculation by Eq. (7), where $\delta\rho_m^{it*1}$ are the values of $\delta\rho_m^c$ at iteration it+1.

$$\delta\rho_m^{it+1} = \delta\rho_m^c + R.\sum_n f_{nm}\cdot(\delta g_n^l - \delta g_n^c) / \sum_n (\alpha_{nm}\cdot f_{nm}) \tag{7}$$

Eq. (7) is intended to calculate all the exact density deviations beginning with inexact density values. The under-relaxation factor R is introduced and the calculations performed iteratively. In the limit in which $\delta g_n^c =$ $\boldsymbol{\delta g}_n^l$, Eq. (6) accepts the solution $\delta\rho_m^c = \delta\rho_m^l$, and Eq. (7) accepts $\lim_{it\to\infty} \delta\rho_m^{it+1} = \delta\rho_m^l$. The $\boldsymbol{\delta g}_n^c$ values in Eq. (7) are updated by means of Eq. (4).

From Eq. (5), Eq.s (7a) can be built to calculate the density deviation corrections from the three Cartesian components K of the gravity deviation corrections.

$$\delta\rho_{Km}^{it+1} = \delta\rho_{Km}^c + R_K \cdot \sum_n f_{Knm} \cdot (\delta g_{Kn}^l - \delta g_{Kn}^c) / \sum_n (\alpha_{Knm} \cdot f_{Knm}) \tag{7a}$$

In the present work no calculations were executed on the basis of eqs. (7a) or their associations other than Eq. (7).

If appropriately chosen, the coefficients $\boldsymbol{f}_{nm}$ will enhance the effect of gravity corrections $\boldsymbol{\delta g}_n^l - \boldsymbol{\delta g}_n^c$ along the directions $\boldsymbol{r_{mn}}$. Moderate directional filtering is advisable in order to substantially include all the *N* gravity vectors in each density correction calculation. The choice of $\boldsymbol{f}_{nm}$ in the direction $r_{mn}$ improves the tendency of $\delta\rho$ corrections in Eq. (6) to be larger in case the gravity vector correction has one direction close to the direction $r_{mn}$. Absolute values of $\boldsymbol{f}_{nm}$ decreasing with $r_{nm}$ will increase the stability of the shallower density values. The variation with distance must insure the gravity data to make smaller corrections at large distance, the same way that density values decline influence on gravity values at long distance. Gravity inversion is so one slow process at large distances in most cases.

The multiplicative coefficients will have here the generic attributed expression of Eq. (8).

$$\boldsymbol{f}_{nm} = \frac{r_{u_{mn}}}{r_{mn}^b} \tag{8}$$

The scalar product of coefficient $\boldsymbol{f}_{nm}$ and gravity correction vector, Eq.s (5, 7), enhances gravity corrections both in the directions close to $\boldsymbol{r_{u_{mn}}}$ and in the smaller distances, acting as one smooth directional and distance filter. The density corrections are improved in those directions at smaller distances.

The values attributed to $\boldsymbol{f}_{nm}$ enter Eq. (7) affecting the gravity corrections at the numerator and the summation in the denominator. At high *b*, the effects on corrections of shallow $\delta\rho$ are greater than at larger $r_{mn}$, so producing quick evolution of shallow $\delta\rho$ and calm down the evolution at far $\delta\rho$. Small *b* is expected to calculate far $\delta\rho$ quicker, possibly unstabilizing shallow calculated $\delta\rho$. The convergence of the iterative process depends on *b* value, here chosen to be *b*=2, equivalent to use $\boldsymbol{f}_{nm} = \boldsymbol{\alpha}_{nm}$.

The utilization of accurate vector gravimetry is needed for the complete procedures in the present version of directionally filtered inversion. Scalar gravity data may possibly be used together with modified versions of the present method, if possible distinguishing between $|\boldsymbol{\delta g}^l|$ and $\delta g_z$.

Zhdanov (2004) use gravity gradient tensor in one inversion method arriving to enhanced resolution. The comparisons of Paoletti (2016) found that inversion data resulting from different components of the gravity gradient tensor or directly from the vertical component of the gravity led to comparable results. Pilkington (2014) found some differences between the results using different tensor components, with preference to combinations of $\boldsymbol{T_{zz}}$ with other components.

Direct measurements or indirect calculations may be used to find vector gravity fields. It is expected that inversion methods using the vector components of gravity together with directional filtering may lead to accurate results. The possibility of alleviating the calculation burden associated to distant density-gravity pairs was not addressed in the present application of the method.

In practical cases in which *N<M* it may be advisable to make nonlinear interpolations and extrapolations of the existing *N* gravity data in order to reach the total number *N=M*. Continuity conditions between local and regional reference fields may be desirable.

## 3. Synthetic models and inversion results

The presented inversion method is applied to two different cases defined in identical geometrical domains.

Volume V, Eq. (2), is a cube with 10km edge, where density fields are defined. It is subdivided in equal parallelepiped volumes δV$_m$, the numbers of nodes in each Cartesian direction being $n_X$= $n_Y$= 5 and $n_Z$= 4. The gravity fields are defined at points located at the upper surface located at z=0. The numbers of representative gravity points are $n_{Xg}$= $n_{Yg}$= 10. The total numbers of *ρ* and *g* points are then *M=N*=100. Two synthetic cases with different density fields are presented. From the chosen density fields a reference field is subtracted, chosen to have the arbitrary value $\rho_{ref} = 3000 kgm^{-3}$.

Eq. (7) is iteratively solved with R automatically varying according to one error parameter based on $|\delta g_n^l - \delta g_n^c|$.

**Case 1**

One quite simple density field is chosen in the V domain, the corresponding deviation field $\delta\rho_m^l = \rho_m^l - \rho_{refm}$ also quantified in Fig. 1. The **z** positions of the centers of the density control volumes δV$_m$ are also shown.

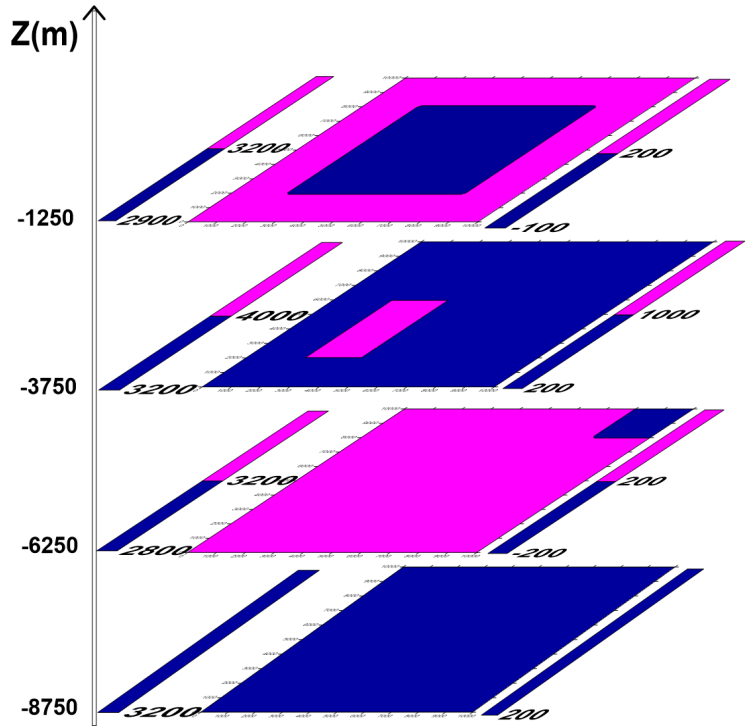

**Fig. 1** Chosen density and density deviation fields (SI) at different depths (case 1).

All the calculations are subsequently performed in terms of the deviation fields *δρ* and *δg*, defined in the previous sections, irrespective of the original density and gravity fields.

From the deviation density field of Fig. 1, the synthetic, here considered exact, deviation gravity $\boldsymbol{\delta g}_n^l$ of Fig. 2 is obtained by means of Eq. (4).

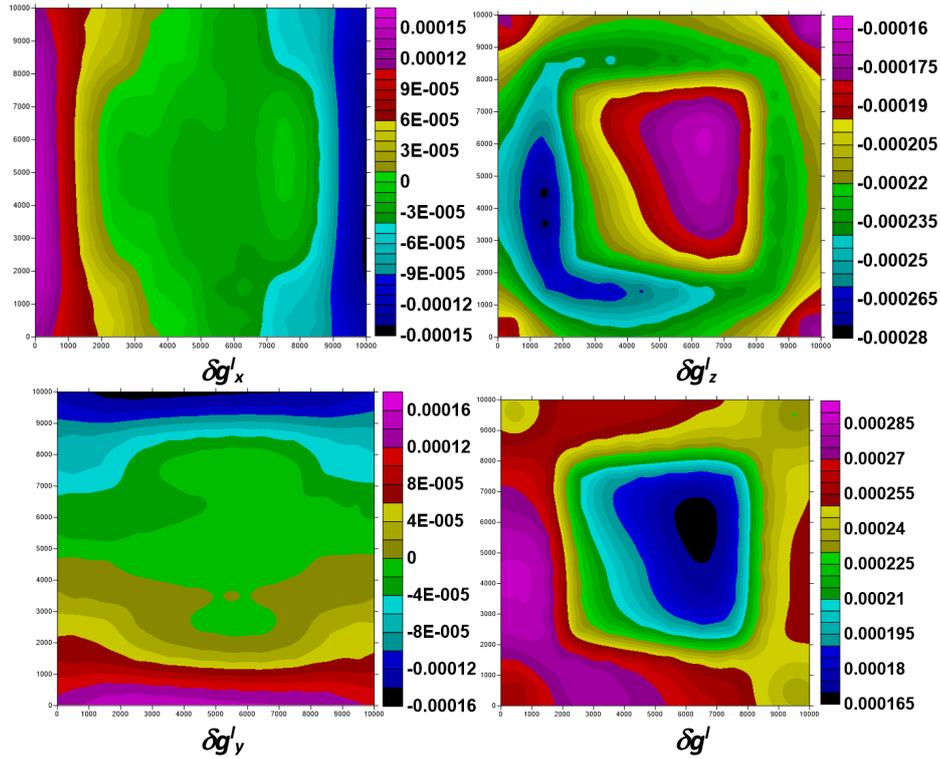

**Fig.2** Maps of the gravity deviation (SI), components and absolute values, calculated by means of Eq. (4) from the original deviation density field of Fig. 1 (case 1).

The deviation gravity field $\delta g_n^l$ of Fig. 2 is subsequently considered to be the original gravity deviation field, from which the deviation density field $\delta \rho_m^c$ is iteratively calculated using Eq. (7) of the present method.

The calculations performed allowed to determine the field $\delta\rho^c(x,y,z)$, the local errors of which, $\delta\rho^c(x,y,z)-\delta\rho^l(x,y,z)$, being shown at Fig. 3 for case 1.

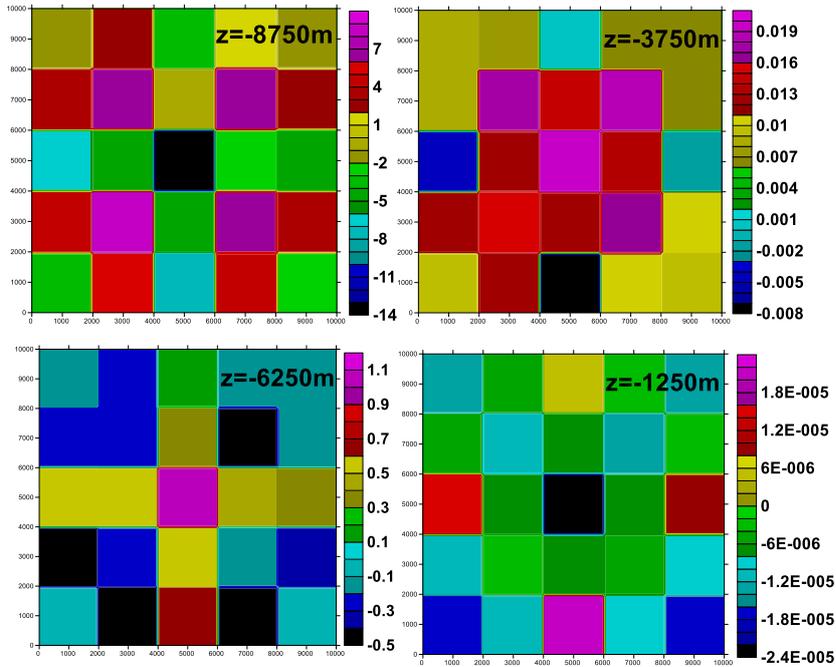

**Fig. 3** Maps of the $\delta\rho_m^c - \delta\rho_m^l$ errors (SI) at different depths, case 1.

The density errors $\delta\rho_m^c - \delta\rho_m^l$, Fig. 3, are reasonably low in case 1, being extremely low at small depths. It can be seen that the calculated density deviations at cells with center on the plane z=-8750m present rather large errors, despite the fact that to this layer uniform density and density deviation have been ascribed.

From the calculated density deviations, the gravity deviation field is now reconstructed by means of Eq. (4). The maps of the calculated relative errors $|\boldsymbol{\delta g_n^c} - \boldsymbol{\delta g_n^l}|/|\boldsymbol{\delta g_n^l}|$ and $(\delta g_n^{cK} - \delta g_n^{lK})/|\delta g_n^{lK}|$ are displayed at fig. 4, the suffix K denoting the three gravity Cartesian components.

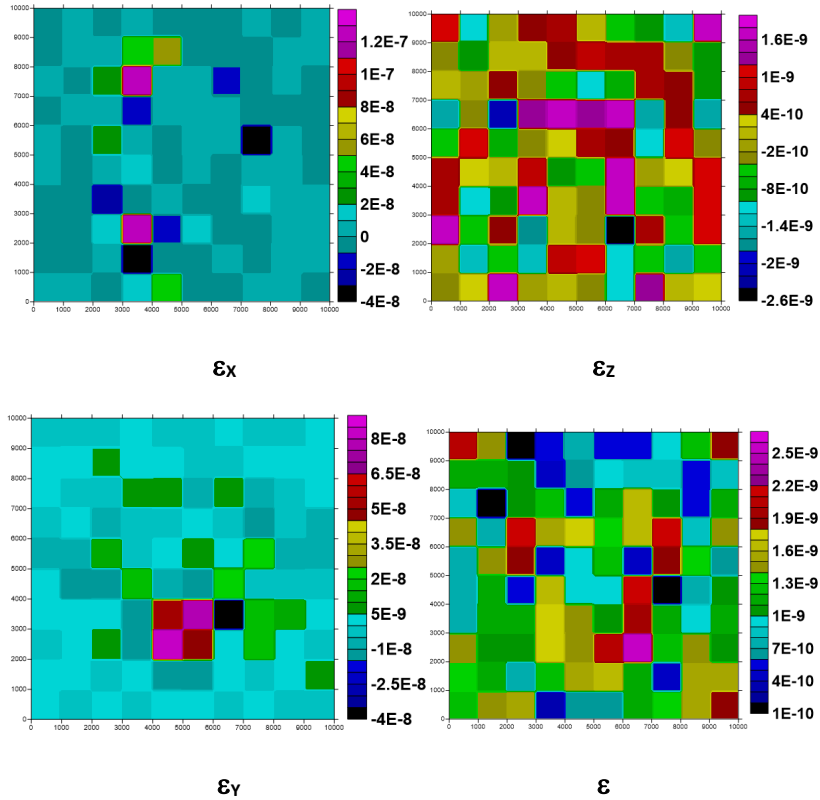

**Fig. 4** Maps of relative gravity deviation errors of components and absolute values (SI), case 1.

All these relative errors are extremely low, once more revealing the small sensitivity of the gravity field relatively to the density at deep layers.

The maps of δg displayed in Fig. 2, presumed exact, are undistinguishable by visual inspection to the calculated ones. This happens even in cases in which the calculated $\delta\rho_m^c$ values have yet considerable differences from the exact ones at lower depths. This strongly suggests that near to correct density values at large depths will be very difficult to reach by the present inversion process.

**Case 2**
The density field of case 2 is represented in Fig. 5. The arbitrated density presents large discontinuities between control volumes at the same level as well as at different levels.

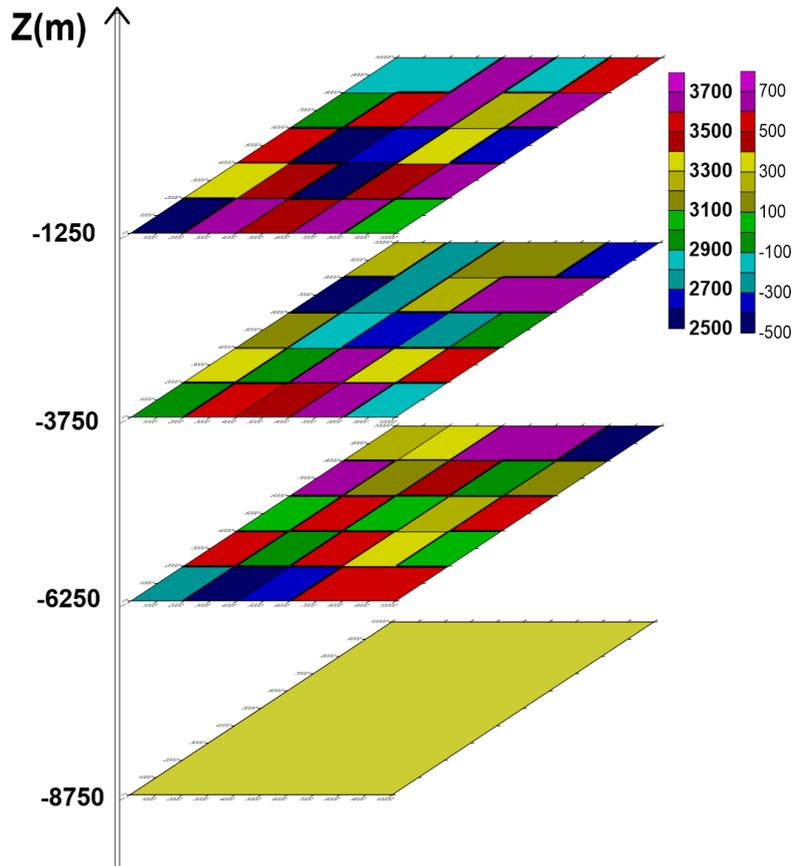

**Fig.** 5 Density field and density deviations (SI) in case 2.

The designated exact gravity deviations obtained from the density field of Fig. 5 are shown at Fig. 6.

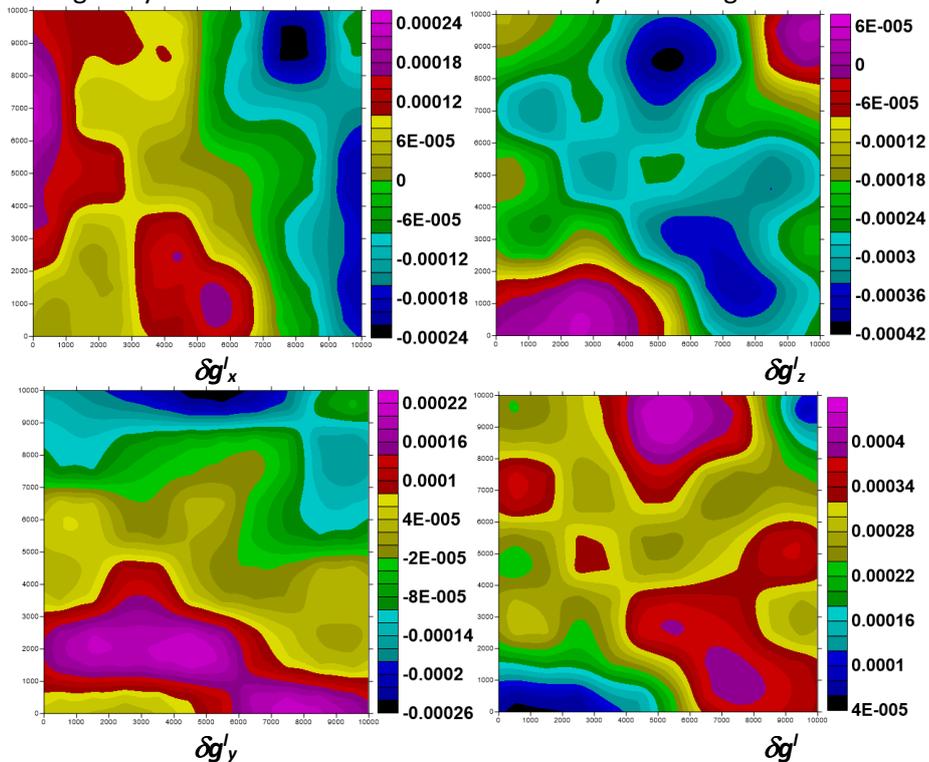

**Fig. 6** Maps of the gravity deviation (SI), components and absolute values, calculated by means of Eq. (4) from the original deviation density field of Fig. 5 (case 2).

The errors of the density deviations calculated by means of the inversion method with respect to the exact density deviations are displayed in Fig. 7.

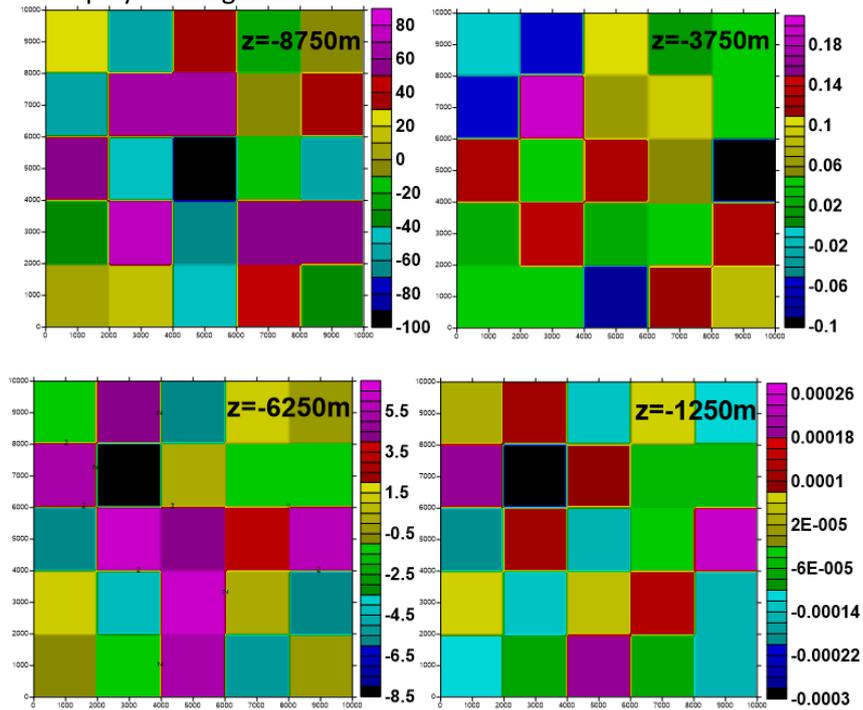

**Fig. 7** Inversion errors (kg/m³) of the density deviation calculated to the exact density deviations at different depths.

Fig. 7 shows the local errors of the calculated density deviation with respect to the exact density deviations in case 2. The extraordinarily precise results at shallow levels, deteriorate as depth increases. At z=-8750m the errors of the density deviations have typical values of tens of kg/m3. The calculation continuation reveals a trend of convergence to the exact values at very slow rates. Likewise case 1, the uniform exact density at the deeper layer is opposed to strong variations of the density values calculated by the inversion method. The large distance to the ground level and the indirect effects of the imprecision in other calculated density in upper layers as well as in the same layer are to be blamed, Eq.s (6,7,4).

The gravity fields obtained from the original density field and obtained by inversion (not independently shown) are indistinguishable by visual inspection. In this way, Fig. 6 may serve as the representation of both the exact gravity field and the gravity field obtained after the density resulting by inversion.

Fig. 8 displays for case 2 the errors of the gravity deviation rebuilt from the calculated density deviation by inversion and the supposed exact gravity deviation, $|\delta g_n^c - \delta g_n^l|/|\delta g_n^l|$.

Despite the considerably large calculated $\delta\rho_m^c - \delta\rho_m^l$ at large depths (Fig. 7), the differences between the calculated and the exact gravity values are extraordinarily small, Fig. 8. With such levels of sensitivity of the gravity field to density at large depths, the inversion procedure generates imprecise densities at profound layers.

Even under the possibility of unique solutions for synthetic cases, practical cases with measurement errors makes the uniqueness hypothesis not directly applicable. Non-unique reference regional fields, as well as different interpolations and extrapolations of the gravity field, imply non-unique results in applied as in synthetic cases.

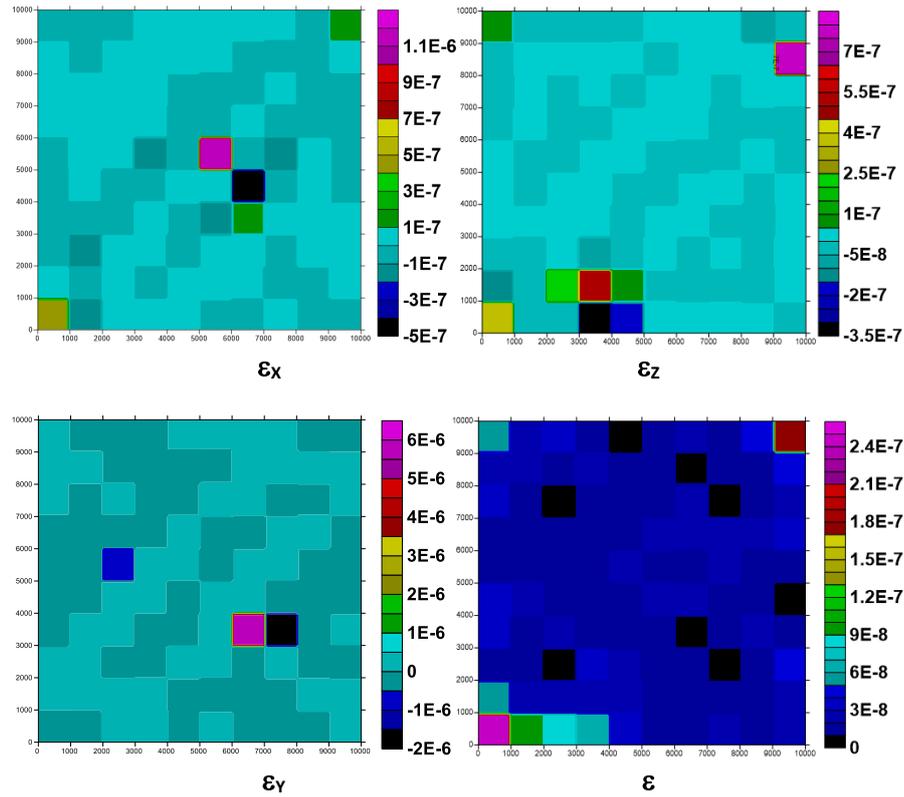

**Fig. 8** Gravity deviation errors, components and absolute values (SI), case 2.

In the course of the present inversion procedures, the exponent of *r* was chosen to have the value *b*=2, with which an adjustable *R* is used that depends on the evolution of the error based on $\left|\boldsymbol{\delta g}_n^l - \boldsymbol{\delta g}_n^c\right|$. Different *b* values were also checked in limited ways. The use of *b*=2 (equivalent to the use of $\boldsymbol{f}_{nm} = \boldsymbol{\alpha}_{nm}$) apparently ensures consistent convergence with no need of further intervention, no study having been followed about convergence rates.

Slow convergence rates are certain at deep layers. The general slow converging rate is more manifest as calculation progresses. Shallower layers exhibit much quicker convergence to the exact results.

Despite the substantial errors of density in deep layers, visual inspection do not reveal differences between the recalculated gravity field and the exact gravity field. This relationship between exact and calculated gravity magnitudes may be seen as one reason for the difficulty of obtaining exact results for density deviations at large depths from gravity data. Small non-zero differences are nevertheless clearly demonstrated in Fig. 8, likewise to what happens in case 1, Fig. 4, even though with better convergence.

For comparison purposes Monte Carlo method has been used to independently solve Eq. (4). The observed overall convergence rates are similar, at least at the relatively small iteration numbers reached in MC calculations.

The role of the directional filtering being reasonably well understood, the degree of decaying with distance of the used filter $\boldsymbol{f}$ has not been adequately studied to determine optimum reduction rates and convergence conditions.

The evolution of the calculated results in the course of the iterative calculations suggests the possibility of exactness at the limit of an infinite number of iterations.

Studying the conditions required to the convergence of the present inversion method may possibly be used as a guide for further developments.

**References**

Camacho, A.G., Fernández, J., and Gottsmann, J., 2011, The 3-D gravity inversion package GROWTH2.0 and its application to Tenerife Island, Spain: Comput Geosci **37**(4), 621–633.

Green, W.R., 1975, Inversion of gravity profiles by use of a Backus-Gilbert approach: Geophysics **40**(5), 763–772.

Li, Y., and Oldenburg, D.W., 1998, 3-D inversion of gravity data: Geophysics, **63**(1), 109–119.

Mosegaard, K., and Tarantola, A., 1995, Monte Carlo sampling of solutions to inverse problems: J. Geophys. Res. **100**(B7), 12431–12447.

Paoletti, V., Fedi, M., Italiano, F., Florio, G., and Ialongo, S., 2016, Inversion of gravity gradient tensor data: does it provide better resolution?: Geophysical Journal International, **205**, 192–202.

Pilkington, M., 2014. Evaluating the utility of gravity gradient tensor components: Geophysics, 79, G1–G14.

Shamsipour, P., Marcotte, D., M. Chouteau, and Keating, P., 2010, 3D stochastic inversion of gravity data using cokriging and cosimulation: Geophysics, **75**(1), 1–10.

Zhdanov, M., Ellis, R., and Mukherjee, S., 2004, Three-dimensional regularized focusing inversion of gravity gradient tensor component data: Geophysics **69**(4), 925–937.